\documentclass[a4paper,11pt]{article}
\pdfoutput=1

\usepackage{jheppub}

\usepackage[T1]{fontenc}

\def\sech{\mathop{\rm sech}\nolimits}

\def\tr{\mathop{\rm tr}\nolimits}

\title{\boldmath The curious case of large-$N$ expansions on a (pseudo)sphere}

\author[a]{Alexander M.~Polyakov}
\author[b]{Zain H.~Saleem}
\author[b]{James Stokes}

\affiliation[a]{Joseph Henry Laboratories, Princeton University, \\
 Princeton, NJ  08544, USA}
\affiliation[b]{Center for Particle Cosmology, Department of Physics and Astronomy,\\University of Pennsylvania, Philadelphia, PA 19104}

\emailAdd{polyakov@princeton.edu}
\emailAdd{stokesj@sas.upenn.edu}
\emailAdd{zains@sas.upenn.edu}

\abstract{
We elucidate the large-$N$ dynamics of one-dimensional sigma models with spherical and hyperbolic target spaces and find a duality between the Lagrange multiplier and the angular momentum. In the hyperbolic model we propose a new class of operators based on the irreducible representations of hyperbolic space. We also uncover unexpected zero modes which lead to the double scaling of the $1/N$ expansion and explore these modes using Gelfand-Dikiy equations.
}

\begin{document} 
\maketitle
\flushbottom

\section{Introduction}
Large-$N$ expansions are important and powerful. They help to understand the behavior of various theories both qualitatively and quantitatively including their phase structure. Most of the known results refer to compact symmetry groups (see however \cite{Niedermaier:2003cq,Niedermaier:2007dw,Friess:2005be,Davis:1982md,Davis:1983rn,Amit:1983eg}). The non-compact case has some interesting features. To reveal and understand them we study in this paper a simple case of AdS hyperbolic sigma models. The target space is defined by the equation
\begin{equation}
	n_\mu n^\mu =  n_0^2-\vec{n}^2  = 1\ .
\end{equation}
This is the Lobachevsky (Euclidean AdS) space. The action is given by
\begin{equation}
	S = \int \left[ (\nabla n_0)^2-(\nabla \vec{n})^2 \right] dt\ .
\end{equation}
This action is strictly positive (despite appearances). It is invariant under the AdS group $SO(1,N)$. The first small puzzle of the model is the following. Consider the 2-point function whose form follows from the symmetry group of AdS,
\begin{equation}
	\langle n_\mu(t_1) n_\nu(t_2) \rangle = -\mathcal{D}(t_1-t_2)\eta_{\mu\nu}\ ,
\end{equation}
where $\eta_{ij}=-\delta_{ij}$ and $\eta_{00}=1$. The negativity of $\langle n_0(t_1) n_0(t_2) \rangle$ clearly contradicts the positivity of norms. The well-known resolution of this puzzle is that the above correlator does not exist, at least naively. The reason lies in the growth of $n_\mu$ which causes the IR divergence of the functional integral. This can easily be seen in the $N=1$ example. Here $n_\mu = (\cosh\theta,\sinh\theta)$ and 
$S \propto \int (\nabla \theta)^2$. As a result,
\begin{align}
	\langle n_\mu(t_1) n^\mu(t_2) \rangle
		& \propto \int \mathcal{D}\theta \, e^{-\int_0^\beta(\nabla\theta)^2 dt} e^{\theta(t_1) - \theta(t_2)}\ ,\\
		& \propto e^{|t_1 -t_2|/\beta}\ ,
\end{align}
which blows up exponentially. An even simpler example is provided by the average $\langle n_0 \rangle = \langle \cosh\theta\rangle$. The symmetry group requires this average to be zero, which is obviously impossible since $n_0 \geq 1$. It is easy to see that the functional integral is divergent and the above average is infinite.

At the same time, if we try to take a sigma model on the sphere and analytically continue it
to the AdS space, the AdS symmetries will be fully respected. Before we clarify the relation between 
the physical and analytically continued correlator we have to establish some facts concerning the compact models.

To understand the dynamics of the two-dimensional sigma model it is helpful to consider a lattice version of it. If we discretize one of the world sheet directions we can write the Hamiltonian of the sphere model as
\begin{equation}
 H = \alpha_0 \sum_x l_x^2 + \frac{1}{\alpha_0} \sum_x ( n_x-n_{x+ 1})^2 \ ,
\end{equation}
where $\alpha_0$ is the coupling and $l_x$ is the angular momentum operator on the $N$-sphere. It is well established that in $1+1$ dimensions the  model develops a mass gap. There are two complementary ways to see this.
The first is the weak coupling limit $ \alpha_0 \to 0$ where we see that the running coupling is given by
\begin{equation}
\alpha(p) = \frac{ \alpha_0}{1- \frac{N-1}{2 \pi} \alpha_0 \log{\frac{\Lambda}{p}}}\ .
\end{equation} 
If the $\beta$-function has no zeros and the model is asymptotically free (negative beta function) we expect that the mass gap is given by
\begin{equation}
 m^2 = \Lambda^2 e^{ -\frac{ 2 \pi}{(N-1)\alpha_0}}\ .
\end{equation}
If we study the model at strong coupling we can see that the second term in the above Hamiltonian can be dropped and the first excitation is given by the $l=1$ mode, corresponding to a mass gap of order $\alpha_0$. This confirms that the beta function has no zeros and the mass gap exists for all values of the coupling.  The lesson of this is that the mass gap of the system depended on the sign of the curvature and the compactness of the model. 

The AdS case differs from the sphere in two important ways: it is non compact and it has a positive beta function as a result of the negative curvature. The absence of a mass gap is corroborated by the energy spectrum which is now given by $ l (l + N-1) $ with the angular momentum having continuous values,
\begin{equation}
	l= -\frac{N-1}{2} + i \rho\ , \quad\quad \rho > 0 \ .
\end{equation}

The two dimensional models in the strong coupling limit can be regarded as a collection of one dimensional rotators. This consideration shows that the understanding of the $1+1$ dimensional sigma model requires as a first step, deeper knowledge of the $0+1$ dimensional sigma model at large $N$. Our goal in this paper is to uncover some useful facts about Lagrange multipliers, angular momenta and saddle points at large $N$. We will make use of the fact that in one dimension there exist two equivalent representations of the path integral. The first approach, due to  E.~Brezin and J.~Zinn-Justin \cite{Brezin:1976qa}, involves a Lagrange multiplier field and is particularly suitable for taking the large-$N$ limit. Our second approach is the Feynman sum over histories method (angular momentum representation) which is applicable for all values of $N$. We will use these path integral representations to study the large-$N$ properties of these models.

The large-$N$ limit is useful when the path integral has a saddle point as $N\to \infty$. This can have very important consequence for correlation functions in the theory. In particular, under certain assumptions, it implies that correlators of singlet operators factorize,
\begin{equation}\label{e:factorization}
	\langle O_{x_1} \cdots O_{x_n} \rangle = \langle O_{x_1} \rangle \cdots \langle O_{x_n} \rangle + \mathcal{O}(1/N)\ .
\end{equation}
As a result, the operators $O_x$ become classical in the $1/N$-expansion. This is similar to what happens in the WKB expansion with $\hbar$ playing the role of $1/N$. We aim to understand to what extent this relation is valid on the $\mathbb{S}^N$ and the AdS$_N$ models.

In section 1 we consider the quantum particle on $\mathbb{S}^N$. We clarify the relationship between the Lagrange multiplier representation and the angular momentum representation by evaluating the integrals by steepest descent. The saddle-point equations match and we thus establish a simple duality between the mass gap and the angular momentum variable. The correlation functions can also be computed on both sides and are shown to agree. On the angular momentum side it is possible to obtain exact formulae for correlators of the invariant singlet operator $z =  \vec{n}(t)\cdot \vec{n}(0)$ as well as correlators of arbitrary polynomials of $z$, which for convenience we choose to be the Gegenbauer polynomials $C_\ell^{d/2}(z)$, where $d = N-1$. Working at both finite and infinite worldline duration $\beta$, we establish the conditions under which \eqref{e:factorization} holds. We find that the number of operator insertions $\ell$ is constrained to be much smaller than $N$. If $\ell$ is of order $N$, then we find that the operator insertions disturb the large-$N$ saddle point and \eqref{e:factorization} fails to hold.

As we move to the AdS$_N$ model with negative curvature we find that
the functional integral is related to the sphere by analytic
continuation. It immediately follows that there are an infinite number
of saddle points from which we select the relevant one. 

We will show
that naive analytic continuation of correlation functions from the
sphere model to AdS$_N$ does not lead to a well defined set of
observables. Our proposal is to instead consider expectations of
generalized Gegenbauer functions which correspond to the complementary
series of irreducible representations of hyperbolic space; namely,
 
\begin{equation}
\langle C_{-d/2+i \rho}^{d/2}(\tilde z) \rangle\ , \quad \quad \tilde
z=n_\mu(t)n^\mu(0) \ .
\end{equation}
 
Unlike polynomials in $\tilde z$, which have infinite expectation
value, our correlators are finite. In contrast to the sphere model,
where large-$N$ factorization followed from the constraint that $\ell
\ll N$, we will see that the saddle point of the functional integral
is necessarily disturbed by the insertion of the Gegenbauer function
and we quantify this disturbance analytically. On the Lagrange
multiplier side of the duality we also discover unexpected zero modes
in the infinite-duration ($\beta \to \infty$) limit. The correlation
functions of these modes satisfy the system of
recursion relations arising in the Gelfand-Dikiy equation.

\section{The sphere model}

\subsection{Lagrange multiplier representation}
We consider quantum particle motion on the $N$-dimensional sphere of unit radius $\mathbb{S}^{N}$. The model is specified by the following partition function for the unit-vector field $\vec{n}$,
\begin{equation}
	Z = \int \mathcal{D}\vec{n} \, \delta(\vec{n}^2 -1) \exp\left[-\frac{1}{2\alpha_0}\int dt \, (\partial_t \vec{n})^2\right]\ .
\end{equation}
The delta function in the above equation restricts the free particle to move on the sphere. We can replace the delta function by imposing the constraint via the Lagrange multiplier instead,

\begin{equation}\label{e:spheredelta}
	\delta(\vec{n}^2 -1) = \int \mathcal{D}\lambda \, e^{-i\tilde{\lambda} (\vec{n}^2-1)}\ ,
\end{equation}
we obtain \cite{Brezin:1976qa}
\begin{align}
	Z 
		& = \int \mathcal{D}\vec{n} \int_{-i\infty}^{+i\infty} \mathcal{D}\lambda \, \exp\left\{-\frac{1}{2\alpha_0}\int dt \left[ (\partial_t \vec{n})^2 + \lambda (\vec{n}^2-1) \right]\right\}\ .
\end{align}
where we have changed variables to $\lambda = i 2\alpha_0\tilde{\lambda}$ and dropped the overall rescaling of the partition function. The exact two-point function is given by
\begin{equation}
	\langle n_i (x) n_j(y) \rangle = \delta_{ij} \alpha_0 \frac{\int \mathcal{D}\lambda \, e^{-W[\lambda]}G(x,y;\lambda)}{\int \mathcal{D}\lambda e^{-W[\lambda]}}\ ,
\end{equation}
where
\begin{equation}
	e^{-W} = \int \mathcal{D} \vec{n} \, e^{-S[\vec{n},\lambda]}, \quad \quad G(x,y;\lambda) = \langle x | (-\nabla^2 + \lambda )^{-1} | y \rangle \ ,
\end{equation}
and
\begin{equation}
W[\lambda]
	= \frac{N}{2}\tr\log(-\partial_t^2 + \lambda) - \frac{1}{2\alpha_0}\int dt \, \lambda(t)\ .
\end{equation}
In the strict $N = \infty$ limit the fluctuations of the Lagrange multiplier can be ignored and it is consistent to set the Lagrange mutliplier equal to a spacetime constant $\lambda = m^2$, the value of which is determined by the saddle point equation; namely, $N\alpha_0 G(x,x;m^2) = 1$. \\

At finite temperature $1/\beta$ the effective action as a function of the constant Lagrange multiplier is
\begin{align}
\frac{\beta}{N}W(\sigma) 
	& = - \frac{1}{2\gamma\beta} \sigma^2 +\frac{1}{2}\sum_{n=-\infty}^{+\infty}\log\left[n^2 + \left(\frac{\sigma}{2\pi}\right)^2\right]\label{e:lambdaaction}\ ,\\
	& = - \frac{1}{2\gamma\beta} \sigma^2 + \log \sinh (\sigma/2)\ ,
\end{align}
where $\gamma = N\alpha_0$ and $\sigma = m\beta$. The two-point function at finite temperature has the following familiar form,
\begin{equation}
	\langle n_i(t) n_j(0) \rangle  
		 = \delta_{ij} \frac{\alpha_0 }{2m} \left[e^{m t} f + e^{-m t} (1+f)\right] \ ,\label{e:spherepropagator}
\end{equation}
where
\begin{equation}\label{bose}
	f = \frac{1}{e^{m\beta} -1}\ , 
\end{equation}
is the usual Bose distribution. The mass gap satisfies $\partial W/\partial m = 0$ or
\begin{equation}
	m = \frac{1}{2}\gamma  \coth \frac{m\beta}{2}\ .
\end{equation}
Using the gap equation we find the following expression for the invariant two-point function,
\begin{equation}\label{e:twopoint}
 \langle \vec{n}(t) \cdot\vec{n}(0) \rangle = \frac{e^{m t}+e^{m(\beta-t)}}{e^{m\beta}+1}\ .
\end{equation}
Taking $\beta \to \infty$ we find the expected zero-temperature mass-gap $m = \gamma/2$.
By analyticitity there are an infinite number of additional saddles with $m^2 < 0$. The negative saddles are approximately given by $m^2 \simeq - \omega_n^2$ ($n \geq 1$) where $\omega_n$ are the Matsubara frequencies. It follows that the negative saddles converge to zero in the $\beta \to \infty$ limit. The selection of the saddle point is governed by the requirement of convergence of the partition function. Recall that the partition function can be represented as a path integral over unconstrained $\vec{n}$ and a Lagrange multiplier field $\lambda$. Taking $\lambda$ to be a spacetime constant, the relevant part of the integral is
\begin{equation}
	\int [d\vec{n}]\exp\left\{-\frac{1}{2\alpha_0}\int d^\mathcal{D} x \left[ (\nabla \vec{n})^2 + \lambda \vec{n}^2 \right] \right\}\ .
\end{equation}
We note at fixed $\lambda$ with $\Re \lambda < 0$ the path integral over $\vec{n}$ is divergent. This implies that $\Re \lambda > 0$ which uniquely fixes the positive saddle. \\

Let us now study the quantum corrections to our system in the Large-$N$ limit. We begin by separating the Lagrange multiplier into a a non-fluctuating component plus a perturbation. The effective action admits the expansion,
\begin{align}\label{e:effaction}
	\frac{\beta}{N}F
		& = - \frac{1}{2\gamma\beta} \sigma^2 + \log \sinh (\sigma/2) + \sum_{n} \Pi(n) \sigma_{n}\sigma_{-n} +    \sum \Gamma(n_1,n_2,n_3)\sigma_{n_1}\sigma_{n_2}\sigma_{-n_1-n_2} + \cdots \ ,
\end{align}
where 
\begin{align}
\Pi(n)
	\propto \sum_k \frac{1}{(n-k)^2 + (\sigma/2\pi)^2}\frac{1}{k^2 + (\sigma/2\pi)^2} \ .
\end{align}
Thus the $1/N$ corrected gap equation is 
\begin{equation}
\frac{1}{2}\frac{ \coth{\sigma}}{\sigma} + \frac{1}{N} \frac{\partial}{\partial \sigma^2} \log \Pi(\sigma^2)+\cdots= \frac{1}{\gamma \beta} \ .
\end{equation}

Canonically normalizing by rescaling $\sigma_n \to \sigma_n /\sqrt{N}$ we find a well-defined expansion in $1/\sqrt{N}$ \cite{Polyakov:1987ez} (in schematic notation),
 \begin{equation}
\langle n_i(x) n_j(0) \rangle = \alpha_0 \delta_{ij} \frac{1}{Z} \int \mathcal{D} \sigma  \exp\left[\sum_{k=0}^\infty N^{-\frac{k}{2}} W_{k+2} \sigma^{k+2} \right] G( x,y , \frac{\sigma}{\sqrt{N}}) \ .
\end{equation}
After taking into account renormalization, this $1/N$ expansion correctly describes the continuum theory in the gapped phase.
\subsection{Angular momentum representation and duality}
Let us now explore the model using exact representation of the partition function in terms of a sum over angular momenta. 
The imaginary-time Feynman path integral (heat kernel) for a particle on $\mathbb{S}^N$ (where $N=d+1$) is \cite{GRSc}
\begin{equation}
	K(\hat{n},\hat{n}';t) 
		= \frac{1}{V_{\mathbb{S}^{d+1}}d}\sum_{\ell=0}^\infty (2\ell + d)C_\ell^{d/2}(\vec{n}\cdot\vec{n}') e^{-\frac{1}{2}\alpha_0 t\ell(\ell + d)}\ .
\end{equation}
It follows that up to multiplicative factors, the thermal partition function (which is the same as the spectral partition function of the Laplace operator) is given by
\begin{equation}
Z =
\sum_{\ell = 0}^\infty
\mathcal{N}(d,\ell)
e^{-\frac{1}{2}\alpha_0 \beta\ell(\ell + d)}, \quad \quad \mathcal{N}(d,\ell) = \frac{(2\ell+d)\Gamma(\ell + d)}{\Gamma(\ell + 1)\Gamma(d)}\ ,
\end{equation}
where $\mathcal{N}(d,\ell)$ is the multiplicity, identically equal to $e^S$ with $S$ being the entropy of a state with given angular momentum.  Defining $f = \ell / d$ and taking $d \to \infty$ we can represent the sum over $\ell$ as an integral over $f$. In the large-$d$ limit we obtain saddle-point equation,
\begin{equation}
 \frac{d}{df} \left[-S(f) + \frac{1}{2}\gamma \beta f(f+1)\right] = 0, \quad S(f) = (f+1)\log(f+1) - f \log f\ ,
\end{equation}
which becomes
\begin{equation}
	\frac{1}{2}(1+ 2f)\gamma \beta = -\log\frac{f}{1+f}\ .
\end{equation}
If we now identify $f$ with the Bose distribution \eqref{bose} then we obtain,
\begin{equation}
	m = \frac{1}{2}\gamma  \coth \frac{m\beta}{2}\ ,
\end{equation}
which is precisely the gap equation obtained from the Lagrange multiplier method. This demonstrates a duality between the angular momentum variable $\ell$ and the Lagrange multiplier $\lambda = m^2$. If we define
\begin{equation}
	\Gamma(\sigma) = \min_f \left[-S(f) + \sigma f\right]\ .
\end{equation}
Then $\sigma = \log\frac{1+f}{f}$ or $f = \frac{1}{e^\sigma - 1}$. This shows that the Lagrange multiplier is the Legendre dual of the angular momentum variable.
\subsection{Large-$N$ factorization}
In this section we will demonstrate the factorization of correlation functions in the large-$N$ limit. 
In terms of the heat kernel the finite temperature correlation function of the Gegenbauer polynomial is given by\footnote{The two-point function $\langle \vec{n} (0) \cdot \vec{n}(t) \rangle$ corresponds to the vector ($\ell = 1$) representation.}
\begin{figure}[h]
\centering
\includegraphics[width=70mm]{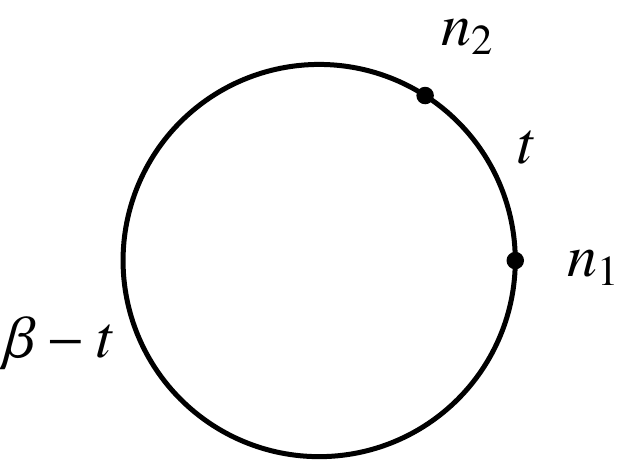}
\caption{Diagram used to define worldline correlators.}
\label{fig:sigmavst}
\end{figure}

\begin{equation}
	\langle C_\ell^{d/2}(\vec{n}(0)\cdot\vec{n}(t)) \rangle = \mathcal{N} \int d \vec{n}_1 d \vec{n}_2 K(\vec{n}_1,\vec{n}_2;\beta-t)K(\vec{n}_1,\vec{n}_2;t) C_\ell^{d/2}(\vec{n}_1\cdot\vec{n}_2)\ ,
\end{equation}
where
\begin{equation}
	\mathcal{N} = \frac{1}{Z} = \frac{1}{\tr \left[K(\vec{n},\vec{n};\beta)\right]} = \frac{1}{\int d\vec{n} K(\vec{n},\vec{n};\beta)} = \frac{1}{V_{\mathbb{S}^{d+1}}K(\vec{n},\vec{n};\beta)}\ .
\end{equation}

Using rotation invariance of the measure we find, 

\begin{align}
	\langle C_\ell^{d/2}(\vec{n}(0)\cdot\vec{n}(t)) \rangle
		& = \frac{V_{\mathbb{S}^{d}}}{K(1;\beta)} \int_{-1}^{+1} dx (1-x^2)^{\frac{d-1}{2}} K(x;\beta-t)K(x;t)C_\ell^{d/2}(x)\ .
\end{align}
It is easier to demonstrate large-$N$ factorization in the zero temperature limit so let us take $\beta \to \infty$ first. This projects out the $\ell = 0$ mode of $K(1;\beta)$ and $K(x;\beta-t)$ giving
\begin{align}
	\langle C_\ell^{d/2}(\vec{n}(0)\cdot\vec{n}(t)) \rangle
		& = V_{\mathbb{S}^{d}}\int_{-1}^{+1} dx (1-x^2)^{\frac{d-1}{2}} K(x;t) C_\ell^{d/2}(x)\ .
\end{align}
Using the completeness relation for Gegenbauer polynomials this becomes,

\begin{equation}
	\langle C_\ell^{d/2}(\vec{n}(0)\cdot\vec{n}(t)\rangle = 
		C_\ell^{d/2}(1)e^{-\frac{\alpha_0}{2}t \ell(\ell+d)}\ . \label{e:gegenbcorr}
\end{equation}
Setting $\ell = 1$ we obtain
\begin{equation}
 \langle \vec{n}(0)\cdot\vec{n}(t) \rangle = e^{-\gamma t /2} \ ,\label{e:nncorr}
\end{equation}
where we have used $C_1^{d/2}(x) = xd$. Recall that this is precisely the propagator computed in the Lagrange multiplier to leading order in $1/N$. 

Let us now consider the large-$N$ limit of this result. Assuming $N \gg \ell$ in \eqref{e:gegenbcorr} we obtain,
\begin{equation}
	\langle C_\ell^{d/2}(\vec{n}(0)\cdot\vec{n}(t))\rangle \simeq C_\ell^{d/2}(1) e^{-(\gamma\ell/2)t} + \mathcal{O}(\ell/N) = C_\ell^{d/2}(1) \langle \vec{n}(0)\cdot\vec{n}(t) \rangle^\ell + \mathcal{O}(\ell/N)\ ,
\end{equation}
 where we have used \eqref{e:nncorr}.
On the other hand, using the asymptotics of Gegenbauer polynomials for $N\gg \ell$ we obtain 
\begin{equation}
  \langle C_\ell^{d/2}(\vec{n}(0)\cdot\vec{n}(t)) \rangle  \simeq C_\ell^{d/2}(1)  \langle ( \vec{n}(0)\cdot\vec{n}(t) )^\ell \rangle \ . \label{e:gegenbcorr2}
\end{equation}
Equating \eqref{e:gegenbcorr} and \eqref{e:gegenbcorr2} we conclude, 
\begin{equation}
	\langle (\vec{n}(0)\cdot\vec{n}(t))^\ell \rangle \simeq \langle \vec{n}(0)\cdot\vec{n}(t) \rangle^\ell + \mathcal{O}(\ell/N)\ ,
\end{equation}
which demonstrates the large-$N$ factorization. It also follows that
\begin{equation}
\langle C_\ell^{d/2}(\vec{n}(0)\cdot\vec{n}(t)) \rangle \sim  C_\ell^{d/2}(\langle\vec{n}(0)\cdot\vec{n}(t)\rangle)  + \mathcal{O}(\ell/N)\ .
\end{equation}

Now we consider the problem at finite $\beta$. Recall the generating function for Gegenbauer polynomials,
\begin{equation}
	\sum_{\ell = 0}C_\ell^{d/2}(z) e^{-\sigma(\ell + d/2)} = e^{-\frac{d}{2}\left[\log 2 + \log(\cosh\sigma - z)\right]}\ .
\end{equation}
Integrating this identity with $e^{\sigma^2/(2\alpha_0 t)}$ we obtain in the large-$d$ limit, 
\begin{equation}\label{e:heatkernel}
	K(z;t) = \int d\sigma e^{-\frac{d}{2}\left[\log(\cosh\sigma - z) - \frac{\sigma^2}{\gamma t}\right]} =  e^{-d W(z,t)}\ .
\end{equation}
The saddle-point values of $\sigma$ (which depends on both $z$ and $t$) solve the equations,
\begin{equation}
	\sigma_t = \frac{\gamma t}{2}\frac{\sinh \sigma_t}{\cosh\sigma_t - z}, \quad \quad \sigma_{\beta-t} = \frac{\gamma (\beta-t)}{2}\frac{\sinh \sigma_{\beta-t}}{\cosh\sigma_{\beta-t} - z}\ .
\end{equation}

The correlation of the Gegenbauer polynomial is given by
\begin{align}
	\langle C_\ell^{d/2}(\vec{n}(0)\cdot\vec{n}(t)) \rangle
	& = \frac{V_{\mathbb{S}^{d}}}{K(1;\beta)} \int_{-1}^{+1} dz (1-z^2)^{\frac{d-1}{2}} K(z;\beta-t)K(z;t)C_\ell^{d/2}(z) \ ,\\
	& = \frac{V_{\mathbb{S}^{d}}}{K(1;\beta)} C_\ell^{d/2}(1)\int_{-1}^{+1} dz e^{d \, \Phi(z)}\ ,
\end{align}
where
\begin{equation}
	\Phi(z) \equiv \frac{1}{2}\log(1-z^2)+(\ell/d)\log z -\left[ W(z,t) + W(z,\beta-t)\right]\ ,
\end{equation}
and we have used $C_\ell^{d/2}(z) \simeq C_\ell^{d/2}(1)z^\ell$ which is valid for $d\gg \ell$. The saddle point satisfies
\begin{equation}
	\Phi'(z_\ast) = \frac{\ell/d}{z_\ast} + \frac{z_\ast}{z_\ast^2 - 1}-\frac{\partial}{\partial z_\ast}\left[W(z_\ast,t) +W(z_\ast,\beta-t)\right]\ .
\end{equation}
Dropping the $\ell/d$ term we obtain,
\begin{equation}
	0 = \frac{z_\ast}{z_\ast^2 - 1} + \frac{1}{2(\cosh\sigma_t - z_\ast)} +  \frac{1}{2(\cosh\sigma_{\beta-t} - z_\ast)}\ ,
\end{equation}
which has the following solution, 
\begin{equation}
	z_\ast = \frac{e^{\sigma_t}+e^{\sigma_{\beta-t}}}{e^{\sigma_t+\sigma_{\beta-t}}+1}\ .
\end{equation}
Substituting into the saddle point equations for $\sigma$ and $\tilde{\sigma}$ we obtain the relationship,
\begin{equation}
	\frac{\sigma_t}{t} = \frac{\sigma_{\beta-t}}{\beta - t}\ .
\end{equation}
This implies that $\sigma_t = m t$ and $\sigma_{\beta-t} = m(\beta - t)$ for some constant $m$. To find $m$ we substitute these ansatze  back into the saddle point equations for $\sigma_t$ and $\sigma_{\beta-t}$ which leads to
\begin{equation}
	m = \frac{1}{2}\gamma \coth \left(\frac{m\beta}{2}\right)\ ,
\end{equation}
which is the correct gap equation. Thus,
\begin{equation}
	z_\ast = \frac{e^{m t}+e^{m(\beta-t)}}{e^{m\beta}+1}\ ,
\end{equation}
which precisely matches the invariant two-point function computed using Lagrange multiplier representation \eqref{e:twopoint}.
Without the insertion of Gegenbauer function we must obtain unity by the normalization condition. Thus the result with the Gegenbauer function included gives
\begin{equation}
	\langle C_\ell^{d/2}(\vec{n}(0)\cdot\vec{n}(t)) \rangle = C_\ell^{d/2}(1) z_\ast^\ell \simeq C_\ell^{d/2}(\langle \vec{n}(0)\cdot\vec{n}(t) \rangle )\ .
\end{equation}
Thus we have shown that if $\ell \ll N$, then large-$N$ factorization holds for finite $\beta$.

It is interesting to understand how the heat kernel \eqref{e:heatkernel} arises from the Lagrange multiplier formalism. It can be obtained by allowing the Matsubara frequencies to become $z$-dependent. More precisely, if we set
\begin{equation}
	\omega_n = \frac{2\pi n + \theta}{\beta}, \quad \quad z = \cos\theta\ .
\end{equation}
in \eqref{e:lambdaaction} then we obtain,
\begin{equation}
	\frac{\beta}{N}W
		= - \frac{1}{2\gamma} m^2 + \frac{1}{2}\log\left(\cosh(m\beta)-z\right)\ .
\end{equation}
which matches the heat kernel result upon setting $\sigma = m\beta$. We see that the expectation value of the Lagrange multiplier depends on $z$. It is easy to check that the expectation value of the Lagrange multiplier is zero for $z < 1 -\gamma_0 \beta/2$. We don't completely understand the significance of this ``phase transition''.
\section{The hyperboloid model}
\subsection{Lagrange multiplier representation: Hidden symmetry and Double scaling}
The $N$-dimensional hyperboloid model is defined in analogy with the sphere model, except that now the $n$-field is taken be a unit-vector in the spacetime $\mathbb{R}^{1,N}$. This leads to two choices depending on whether we take the positively-curved, single-sheeted hyperboloid (de Sitter space) or the negatively-curved, two-sheeted hyperboloid (hyperbolic space).  The effective action is obtained from the sphere model by taking $\gamma \to -\gamma$ and $\sigma \to i\sigma$,
\begin{align}\label{e:lambdaaction}
\frac{\beta}{N}W(\sigma) 
	& =  -\frac{1}{2\gamma\beta} \tilde{\sigma}^2 +\frac{1}{2}\sum_{n=-\infty}^{+\infty}\log\left[n^2 - \left(\frac{\tilde{\sigma}}{2\pi}\right)^2\right] \ , \\
	& =  -\frac{1}{2\gamma\beta} \tilde{\sigma}^2 + \log \sin (\tilde{\sigma}/2) \ ,
\end{align}
where $\tilde{\sigma}=\tilde{m}\beta$ and  $\lambda=-\tilde{m}^2$ is the background value of the Lagrange multiplier which satisfies the gap equation $\partial W/\partial\tilde{m} = 0$ or
\begin{equation}
	\tilde{m} = \frac{1}{2}\gamma  \cot \frac{\tilde{m}\beta}{2} \ ,
\end{equation}
and the naive invariant two-point function obtained from analytic continuation from the sphere is 
\begin{equation}
	\langle n^\mu (t) n_\mu(0) \rangle = -\frac{\gamma}{2\tilde{m}}\frac{\cos \tilde{m}(t-\beta/2)}{\sin(\beta\tilde{m}/2)} =  -\frac{\cos \tilde m \left(t-\beta/2\right)}{\cos(\tilde m \beta/2)}\ .
\end{equation}
The gap equation no longer has solutions with $\lambda >0$ but there are infinitely many solutions with $\lambda < 0$.
For large $\beta$ the solutions can be approximated by
\begin{equation}
	\tilde{\sigma} = k\pi\ , \quad \quad k = 1,2,\ldots \ .
\end{equation}
Hence the imaginary mass gap $\tilde{m} = k\pi/\beta$ decays with system size. Considering only the $k=1$ solution we find that when $\beta \to \infty$ we have $\sigma = \pi$. Expanding away from zero temperature by writing $\sigma = \pi + \epsilon$ we obtain the following dependence of $\epsilon$ on temperature to first order in $1/(\gamma\beta)$,
\begin{equation}
	\epsilon = -\frac{2}{\gamma \beta}\ .
\end{equation}

The effective action admits a similar expansion of the form \eqref{e:effaction} except
the correlation functions exhibit strikingly different behavior in the $\beta \to \infty$ limit. Let us consider the two-point function
\begin{align}
\Pi(n)
	& = \int_0^\beta dt e^{i\omega_n t} \langle n^\mu (t) n_\mu(0) \rangle^2, \quad\quad \omega_n= \frac{2 \pi n}{\beta} \ .
\end{align}
In the limit $\sigma \to \pi$ we obtain
\begin{equation}
	\langle n^\mu (t) n_\mu(0) \rangle =  -\left|\frac{\beta}{2\pi}\sin(\omega_1 t/2)\right| \ ,
\end{equation}
and thus
\begin{align}
	\Pi(n)
		& = -\frac{\beta^2}{16\pi^2} \int_0^\beta dt \left[e^{i(\omega_n - \omega_{-1})t}+ e^{i(\omega_n - \omega_{1})t}-2e^{i(\omega_n - \omega_{0})t}\right] 
		\propto \delta_{n,0} - \frac{1}{2}(\delta_{n,1}+\delta_{n,-1}) \ .
\end{align}
We therefore find the curious result that the polarization operator vanishes for all $n \neq 0,\pm 1$. Remarkably, this behavior generalizes to all orders in the $1/N$ expansion. Let us check this explicitly for the three-point function,
\begin{align}
	& \Gamma(n_1,n_2,-n_1-n_2) \notag \\
		& =\frac{1}{4(\omega_{n_1}+\omega_{n_2})}\int_0^\beta d \tau_1 \biggl[\frac{\cos(\omega_{n_1}+\omega_1)\tau_1-\cos(\omega_{n_1}\tau_1)}{\omega_{n_1}+\omega_{n_2} +\omega_1}+\frac{\cos(\omega_{n_1}+\omega_{-1})\tau_1-\cos (\omega_{n_1}\tau_1)}{\omega_{n_1}+\omega_{n_2} + \omega_{-1}} + \notag \\
		& \quad +(n_1\leftrightarrow n_2)\biggr] \ .
\end{align}
Notice that if either $n_1$ or $n_2$ vanishes or equals $\pm 1$ then we obtain a finite contribution. If, on the other hand neither $n_1$ nor $n_2$ equals $0,\pm 1$ then $\Gamma$ vanishes. Any potential divergences cancel.
 This pattern continues for all correlation functions $W_k(n_1,\ldots n_{k-1},n_k;\mu)$. That is,
\begin{equation}
	W_k(n_1,\ldots n_{k-1},n_k;\mu) = \begin{cases} \mathcal{O}(\epsilon) & \textrm{if all} \quad |n_i| > 1 \\
	\mathcal{O}(1) & \textrm{otherwise}  \end{cases} \ ,
\end{equation}
where $n_k = -n_1 - \cdots - n_{k-1}$.
\begin{figure}[h]
\centering
\includegraphics[width=50mm]{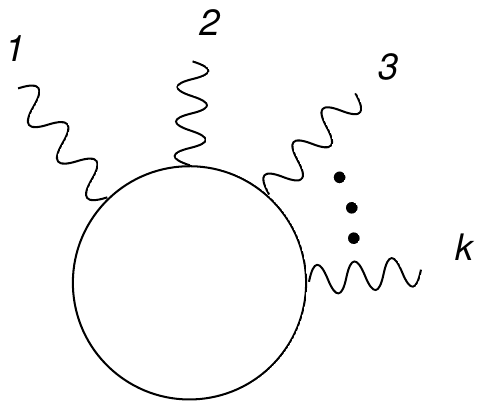}
\caption{Feynman diagram for $W_k$. The solid line is the $n$-field and wavy lines represent the Lagrange multiplier field. Special cases are $\Pi = W_2$ and $\Gamma = W_3$.}
\end{figure}

The vanishing of correlation functions suggests an alternative expansion parameter in the infinite-duration ($\beta \to\infty$) limit, much like the double-scaling limit in matrix models. If we consider the modes in the effective action for which $W_k \sim \mathcal{O}(\epsilon)$, then we can write $W_k = \epsilon w_k$ so that schematically
\begin{equation}
\langle n_i(x) n_j(y) \rangle = \alpha_0 \delta_{ij} \frac{1}{Z} \int \mathcal{D}\sigma \, e^{(N\epsilon w_2 \sigma^2 + N\epsilon w_3  \sigma^3 + N\epsilon w_4 \sigma^4 + \cdots)} G( x,y,\sigma) \ .
\end{equation}
Rescaling $\sigma_n \to \sigma_n/\sqrt{N\epsilon}$ we find
\begin{equation}
\langle n_i(x) n_j(0) \rangle = \alpha_0 \delta_{ij} \frac{1}{Z} \int \mathcal{D}\sigma  \exp\left[\sum_{k=0}^\infty (N\epsilon)^{-\frac{k}{2}} w_{k+2}  \sigma^{k+2} \right] G( x,y , \frac{\sigma}{\sqrt{N\epsilon}}) \ .
\end{equation}
Thus perturbation theory is well defined for these modes with the expansion parameter $1/\sqrt{N\epsilon} \sim \sqrt{\beta/N}$.
\subsection{Relationship with the Gelfand-Dikiy equation and Korteweg-de Vries symmetry}
There is an interesting relationship between the correlation functions in the double-scaling limit and the recurrence relations for the Gelfand-Dikiy equation\footnote{For related work on the Gelfand-Dikiy equations in the cotext of $O(N)$ models see \cite{Feinberg:1994qq,Feinberg:1994fq}.}. We will show that the variational derivative $\delta W/\delta\sigma(t) = g(t)$ is a solution of the Dikii-Gelfand equation in the potential $u(t)$ proportional to $\sigma(t)$. In terms of Fourier modes, our main result is that the different orders of the solution of the Gelfand-Dikiy equation are related to the correlation functions $W_k$ by the relation\footnote{In this section we use the normalization $\beta = 2\pi$.}
\begin{equation}
g_n^{(k)} \propto \sum_{n_i} W_{k+1}(n,n_1,\ldots n_{k}) u_{n_1} \cdots  u_{n_k} \delta_{n+n_1+\cdots + n_{k},0} \ .
\end{equation}
Recall that the Gelfand-Dikiy equation for the coincidence limit $g(t) = G(t,t)$ of the Green function $G(t,t')$ in a potential $u(t)$ is given by
\begin{equation}\label{e:DKequation}
	g''' + g' = 2u' g + 4 u g' \ .
\end{equation}
Expanding the Green function as $g = \sum_{j=1}^\infty g^{(j)}$ the resulting set of recurrence relations are
\begin{align}
	\partial_t^3 g^{(1)} + \partial_t g^{(1)}
		& = 0 \ , \\
	\partial_t^3g^{(j+1)} + \partial_t g^{(j+1)}
		& = 2 \partial_t u \, g^{(j)} + 4 u \partial_t g^{(j)}, \quad \quad j \geq 1 \ .
\end{align}
In Fourier space the first relation is $n(n^2-1)g^{(1)}_n = 0$ which implies
\begin{equation}
	g^{(1)}_n = \left[\delta_{n,0}  -\frac{1}{2}(\delta_{n,1}+\delta_{n,-1})\right]u_n \ .
\end{equation}
The coefficients in this equation are fixed by comparing with $\Pi(n)$. Similarly the second order relation in Fourier space is
\begin{align}
	n(n^2-1)g^{(2)}_n
		& = - 2 \sum_m u_{n-m} (n+m)g^{(1)}_m \ , \\
		& = n(n^2-1)g^{(2)}_n= - \left[2 n u_0 u_n - ( (n+1)u_{n-1} u_1 + (n-1)u_{n+1} u_{-1}) \right]\ .
\end{align}
where we have substituted the expression for $g^{(1)}_n$ in the second line. This equation fixes $g^{(2)}_{n}$ except for $n = 0,\pm1$. The coefficients of $\delta_{n,0}$, $\delta_{n,1}$ and $\delta_{n,-1}$ are determined by the sum rules which follow from setting $n=0,\pm 1$ in the third order equation; namely,
\begin{align}
	\sum_m u_{-m} g^{(2)}_m m
		& = 0 \ , \\
	\sum_m u_{1-m} g^{(2)}_m (m+1)
		& = 0 \ , \\
	\sum_m u_{-1-m} g^{(2)}_m (m-1)
		& = 0 \ .
\end{align}
Substituting the general expression for $g^{(2)}_n$ we obtain the system of equations
\begin{align}
\sum_n' g^{(2)}_n u_{-m}m + g^{(2)}_{-1} u_{-1} - g^{(2)}_{-1} u_1 
	& = 0 \ , \\
\sum_n' g^{(2)}_n u_{1-m}(m+1) + g^{(2)}_{1} u_{1} + 2  g^{(2)}_{0} u_0 
	& = 0 \ , \\
\sum_n' g^{(2)}_n u_{-1-m}(m-1) - g^{(2)}_{-1} u_{-1} - 2 g^{(2)}_{0} u_0 
	& = 0 \ ,
\end{align}
where the prime means we exclude $m=0,\pm1$ in the sum. Using the fact that $\sum_n' g^{(2)}_n u_{-m}m = 0$ and $u_{n} = u_{-n}$ we obtain  the solution
\begin{equation}
	g^{(2)}_1 = g^{(2)}_{-1} = -\left[\frac{1}{2u_0} u_1 g^{(2)}_0 + \sum_n' g^{(2)}_n u_{1-m}(m+1) \right] \ .
\end{equation}
The statement of the previous section follows from the fact that in the case when $u_{0,\pm 1}$ is zero, iterations of the Dikiy-Gelfand equation give zero in all orders.  In other words, there is no interaction without zero modes $u_{0,\pm 1}$.

Another interesting feature of $W[u]$ is its invariance under Korteweg-de Vries evolution. Multiplying \eqref{e:DKequation} by $u$ and integrating by parts we get
\begin{equation}
	\int dt \left(\frac{\delta W}{\delta u}\right)\delta u = 0 \ ,
\end{equation}
if $\delta u(t) = \epsilon (u'''-6uu'+u')$. The implications of this symmetry of the Lagrange multiplier are still to be understood.
\subsection{Angular momentum representation and duality}
The imaginary-time Feynman path integral (heat kernel) for a particle moving on the hyperboloid is \cite{GRSc}

\begin{equation}
	K(n_1,n_2;t) = \frac{2i}{V_{\mathbb{S}^{d+1}}d} \int_0^\infty d\rho \, \rho \, C^{d/2}_{-d/2 +i\rho}(n_{1}^\mu n_{2\mu}) e^{-\frac{1}{2}\alpha_0 t\left[\rho^2 + d^2/4\right]} \ ,
\end{equation}
which is precisely the analytic continuation from the sphere with $\ell = -d/2 + i \rho$. To obtain the partition function we consider the coincidence limit which gives
\begin{equation}
	Z = \int_{0}^\infty d\rho \, \mathcal{N}(d,\rho)e^{-\frac{1}{2}\alpha_0\beta\left[\rho^2 + d^2/4\right]} \ .
\end{equation}
Using gamma function identities we find
\begin{align}
	C_{-d/2+i \rho}^{d/2}(1)
		& = i\frac{\Gamma^2(d/2)}{\Gamma(d)}   \rho \exp \left[\sum_{n=1}^{d/2-1}\log\left(1+\frac{\rho^2}{n^2}\right)\right] \ .
\end{align}
Defining $v = \rho /d$ we obtain
\begin{align}
	\sum_{n=1}^{d/2-1}\log\left(1+\frac{\rho^2}{n^2}\right)
		& = d\left[\sum_{n=1}^{d/2-1} \frac{1}{d} \log\left(1+\frac{v^2}{(n/d)^2}\right)\right] \ ,
\end{align}
which becomes $d\int_{0}^{1/2}dx \log(1+v^2/x^2)$ in the large-$N$ limit.
The saddle point equation in $v$ is
\begin{equation}
	\frac{d}{dv}\left[ \log \mathcal{N} - \beta v^2 d^2 \right] = 0 \ ,
\end{equation}
which implies
\begin{equation}
	v = \frac{1}{2}\cot\left(\frac{\gamma\beta v}{2}\right) \ . \label{e:lobachevskygap}
\end{equation}
This is related to the gap equation obtained from the Lagrange multiplier representation by
\begin{equation}
	\tilde{m} = \gamma v \ .
\end{equation}
We see that there is again a simple duality between $\rho$ and the Lagrange multiplier.

\subsection{Large-$N$ factorization}
The heat kernel on the hyperboloid in the large-$N$ limit is given by
\begin{equation}
	K(z;t) \sim \int d\tilde{\sigma} \, e^{-\frac{d}{2}\left[\log(\cos\tilde\sigma - z)-\frac{\tilde\sigma^2}{\gamma t}\right]} = e^{-W(z,t)} \ .
\end{equation}
Setting $z=1$ and varying with respect to $\tilde\sigma$ we find the expected relation $\tilde\sigma = \tilde{m} t$ where $\tilde{m}$ satisfies the Lobachevsky gap equation. The situation is very different when we consider the correlator of the Gegenbauer function on AdS. Inserting the Gegenbauer function into the path integral gives
\begin{align}
\langle C_{-d/2+i\rho}^{d/2}(n^\mu(0) n_{\mu}(t)) \rangle
		& = \mathcal{N}\int dn_1 dn_2 K(n_1, n_2;\beta - t)K(n_1,n_2;t)C_{-d/2+i\rho}^{d/2}(n_1^\mu n_{2\mu}) \ , \notag \\
		& = \frac{V_{\mathbb{S}^{d}}}{K(1;\beta)}\int_{1}^\infty dz (z^2 - 1)^{\frac{d-1}{2}}K(z;\beta - t)K(z;t)C_{-d/2+i\rho}^{d/2}(z) \ .
\end{align}
The saddle-point values for $\tilde\sigma$ are now given by
\begin{equation}
	\tilde\sigma_t = \frac{\gamma t}{2}\frac{\sin\tilde\sigma_t}{z-\cos\tilde\sigma_t}, \quad \quad \tilde\sigma_{\beta-t} = \frac{\gamma (\beta -t )}{2}\frac{\sin\tilde\sigma_{\beta-t}}{z-\cos\tilde\sigma_{\beta-t}} \ .
\end{equation}
These expressions coincide with the analytic continuation of the corresponding equations on the sphere.
The saddle-point in $z$ is determined by the method of steepest descent on the following integral,
\begin{align}
	\langle C_{-d/2+i\rho}^{d/2}(n^\mu(0) n_{\mu}(t)) \rangle
		& = \frac{V_{\mathbb{S}^d}}{K(1;\beta)} C_{-d/2+i\rho}^{d/2}(1)\int_1^\infty dz e^{d \Phi(z)} \ ,
\end{align}
where
\begin{equation}
	\Phi(z) = \frac{1}{2}\log(z^2-1) + \frac{1}{2}\log\left(\frac{2}{1+z}\right) - W(z,t) - W(z,\beta-t)\ .
\end{equation}
Here we have used an important property of the Gegenbauer functions in the large-$d$ limit; namely,
\footnote{The asymptotics are obtained using \begin{align}
	C_{-d/2+i\rho}^{d/2}(\cosh r)
		& = C_{-d/2+i\rho}^{d/2}(1) {_2F_1}\left(1/2-i\rho,1/2+i\rho;\frac{d+1}{2};\frac{1-\cosh r}{2}\right)\sech^{d-1}(r/2) \ .
\end{align}},
\begin{equation}
C_{-d/2+i\rho}^{d/2}(\cosh r) \sim C_{-d/2+i\rho}^{d/2}(1)\sech^{d-1}(r/2) \ .
\end{equation}
Hence we conclude that the saddle-point equation of $z$  is \emph{not} the analytic continuation of the sphere result due to the presence of the second term, which leads to the disturbed saddle point equation,
\begin{equation}
	0=\Phi'(z_\ast) = \frac{z_\ast}{z_\ast^2 - 1}-\frac{1}{2(z_\ast+1)} - \frac{\partial}{\partial z_\ast}\left[W(z_\ast,t) +W(z_\ast,\beta-t)\right] \ ,
\end{equation}
which is solved by
\begin{equation}
	z_\ast = 1 + 2\sin(\tilde\sigma_t/2)\sin(\tilde{\sigma}_{\beta-t}/2) \ .
\end{equation}
The system of saddle point equations for $z$, $\tilde\sigma_t$ and $\tilde\sigma_{\beta-t}$ can be solved simultaneously as on the sphere, but does not yield a simple analytical result,
\begin{equation}
	\frac{\tilde\sigma_t}{\tilde\sigma_{\beta-t}} =  \frac{t}{\beta -t} \frac{\cos(\tilde\sigma_t/2)}{\cos(\tilde\sigma_{\beta-t}/2)} \ .
\end{equation} 
Thus we reach the conclusion that insertion of the Gegenbauer function significantly disturbs the $N=\infty$ saddle point of the functional integral. Solving the system without the crucial disturbance term $\frac{1}{2}\log\left(\frac{2}{1+z}\right)$ leads to the erroneous conclusion that $z_\ast$ is related to the sphere by a mere analytic continuation, 
\begin{equation}\label{e:undisturbed}
z_\ast= \frac{\cos\left( \frac{\tilde\sigma_t-\tilde\sigma_{\beta-t}}{2} \right)}{\cos\left( \frac{\tilde\sigma_t+\tilde\sigma_{\beta-t}}{2}\right)} \ .
\end{equation}
where $\tilde\sigma_t = \tilde{m}t$ and $\tilde\sigma_{\beta-t} = \tilde m (\beta -t)$.

\begin{figure}[h]
\centering
\includegraphics[width=70mm]{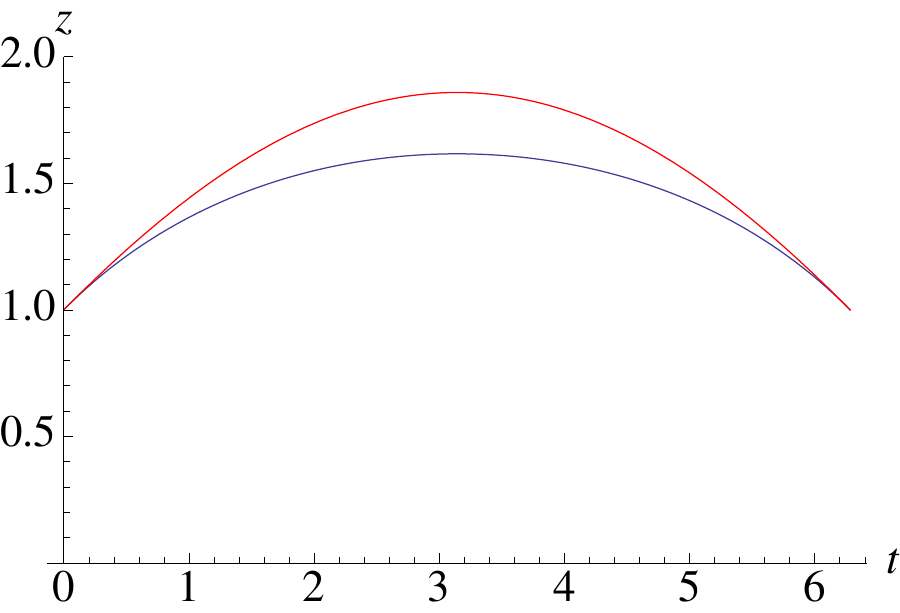}
\caption{The disturbed $z_\ast$ saddle point (blue) as a function of $t$ compared to the undisturbed result (red) for $\beta = 2\pi$ and $\gamma =1 $.}
\end{figure}

\section{Discussion and Future Directions}
We have demonstrated that a duality exists between the Lagrange multiplier and angular momentum. In fact they are related by Legendre transformation. We have shown that large-$N$ factorization holds in the $\mathbb{S}^N$ model provided that the number of operator insertions is much smaller than the dimension of the sphere.

In contrast the AdS$_N$ sigma model does not exhibit large-$N$ factorization under any circumstances because the insertion of an operator necessarily disturbs the saddle-point of the functional integral. We found analytical expressions for the system of disturbed saddle point equations. In addition we found unexpected zero modes in the $\beta \to \infty$ limit. It will be interesting to understand the significance of these modes in $1+1$ dimensional sigma model which can be considered as a collection of rotators with interactions. It will also be interesting to further explore the relationship of these modes with the Gelfand-Dikiy equation.

Another potentially fruitful direction are the models with negative curvature and a compact target space. These can be obtained from quotients of the form AdS${_N}/\Gamma$ where $\Gamma$ is a discrete subgroup of $SO(1,N)$. Such spaces generally have non-trivial homotopy and thus we can expect to see topological phase transitions \cite{gubser} and perhaps second-order phase transitions too \cite{Polyakov:2007mm}.

Perhaps the most important future direction is the de Sitter sigma model. Perturbation theory in $\alpha_0$ is identical to the sphere because in any number of loops the contribution to the beta function is expressed solely in terms of powers of the scalar curvature. However the continuous spectrum due to non compactness suggests that non-perturbatively it should differ. It is tempting to speculate
that the negative saddles in the sphere model play a role in this non-perturbative regime. It will  also be interesting to see if there
exists a symmetry in the de Sitter sigma model much like in the AdS case.

\acknowledgments
Z.S. and J.S. would like to thank Yi-Zen Chu, Loyal Durand, Steven Gubser, Randall Kamien, Max Niedermaier, Hernan Piragua  and Edward Witten for useful discussions and correspondence. J.S. is supported in part by NASA ATP grant NNX14AH53G. Z.S. is supported in part by DOE Grant DOE-EY-76-02- 3071. The work of A.P. was partially supported by the US NSF grant number PHY-1314198.

\end{document}